\documentclass[10pt]{proc}

\usepackage{latexsym, amsmath, amsfonts, amssymb, amsthm, graphicx, epsfig, array, longtable, calc, multirow, hhline, verbatim, color, hyperref, authblk, enumerate} 
\usepackage[square,sort,comma,numbers]{natbib}

\def\b{\mathbf b}
\def\e{\mathbf e}
\def\f{\mathbf f}
\def\h{\mathbf h}
\def\v{\mathbf v}
\def\x{\mathbf x}

\def\A{\mathbf A}
\def\C{\mathbf C}
\def\G{\mathbf G}
\def\H{\mathbf H}
\def\I{\mathbf I}
\def\J{\mathbf J}
\def\L{\mathbf L}
\def\M{\mathbf M}
\def\N{\mathbf N}
\def\SS{\mathbf S}
\def\Q{\mathbf Q}
\def\V{\mathbf V}
\def\X{\mathfrak X}

\def\1{\mathbf 1}
\def\0{\mathbf 0}

\def\NN{\mathbb N}
\def\RR{\mathbb R}

\def\tr{\mathrm{tr}}
\def\vech{\mathrm{vech}}

\title{Ascent with Quadratic Assistance\\ for the Construction of Exact Experimental Designs}
\author{Lenka Filov\'{a}\thanks{Faculty of Mathematics, Physics and Informatics, Comenius University in Bratislava, Slovakia\\ email: filova@fmph.uniba.sk}, Radoslav Harman\thanks{Faculty of Mathematics, Physics and Informatics, Comenius University in Bratislava, Slovakia and Johannes Kepler University Linz, Austria\\ email: harman@fmph.uniba.sk}}
\date{}

\begin{document}
	\maketitle
	
\begin{abstract}
In the area of statistical planning, there is a large body of theoretical knowledge and computational experience concerning so-called optimal approximate designs of experiments. However, for an approximate design to be executed in practice, it must be converted into an exact, i.e., integer, design, which is usually done via rounding procedures. Although rapid, rounding procedures have many drawbacks; in particular, they often yield worse exact designs than heuristics that do not require approximate designs at all.
		
In this paper, we build on an alternative principle of utilizing optimal approximate designs for the computation of optimal, or nearly-optimal, exact designs. The principle, which we call ascent with quadratic assistance (AQuA), is an integer programming method based on the quadratic approximation of the design criterion in the neighborhood of the optimal approximate information matrix.
		
To this end, we present quadratic approximations of all Kiefer's criteria with an integer parameter, including $D$- and $A$-optimality and, by a model transformation, $I$-optimality. Importantly, we prove a low-rank property of the associated quadratic forms, which enables us to apply AQuA to large design spaces, for example via mixed integer conic quadratic solvers. We numerically demonstrate the robustness and superior performance of the proposed method for models under various types of constraints. More precisely, we compute optimal size-constrained exact designs for the model of spring-balance weighing, and optimal symmetric marginally restricted exact designs for the Scheff\`{e} mixture model. We also show how can iterative application of AQuA be used for a stratified information-based subsampling of large datasets under a lower bound on the quality and an upper bound on the cost of the subsample. 
		
		
\end{abstract}

\section{Introduction}

Suppose that we intend to perform an experiment consisting of a set\footnote{Note that we implicitly assume that reordering of the trials does not influence the relevant properties of the experimental design.} of trials. Assume that the observed response in each trial depends on a design point chosen from a finite design space $\X=\{\x_1,\ldots,\x_n\}$. For instance, $\X$ may be the set of all available combinations of levels of several discrete factors\footnote{In some experimental situations, the set of available design points can be modeled as a continuous domain. However, in many applications, the design space is finite. This is the case if each factor has - in principle or effectively - only a finite number of levels that the experimenter can select, or if the optimal design problem corresponds to data sub-selection (see the examples in Section \ref{sec:EX}). Moreover, the method proposed in this paper can also be useful for solving the problems with continuous design spaces; cf. Section \ref{Sec:misc}.}. 
\bigskip

An ``exact'' design (ED) is a selection $\xi$ of design points, not necessarily distinct, to be used for individual trials. We will formalize an ED $\xi$ as a non-negative integer-valued vector $(\xi_1,\ldots,\xi_n)^T \in \mathbb{N}_0^n$,\footnote{The symbols $\RR$, $\RR_+$, $\NN$, $\NN_0$, and $\RR^{k \times n}$ denote the sets of real, non-negative real, natural, non-negative integer numbers, and the set of all $k \times n$ real matrices, respectively.} where $\xi_i$, called the $i$-th weight, represents the number of trials to be performed at the design point $\x_i$, $i=1,\ldots,n$.\footnote{Therefore, we do not represent designs by normalized (probability) measures, as is frequently done in optimal design, but by non-normalized vectors of numbers of trials.} An ``approximate'' design (AD), $\xi=(\xi_1,\ldots,\xi_n)^T \in \RR_+^n$, is allowed to have general non-negative components, which means that the weight $\xi_i$ is a continuous relaxation of the integer number of trials to be performed at $\x_i$, $i=1,\ldots,n$.\footnote{Approximate designs are sometimes also called ``continuous'' designs, which refers to the continuity of the space of designs, not the design space.} Thus, an AD must be converted into an ED prior to its application in a real experiment. 
\bigskip

Let $\Xi^E_{\A,\b}=\{\xi \in \NN_0^n: \A \xi \leq \b\}$ be a non-empty set of permissible EDs, where $\A \in \RR^{k \times n}$ and $\b \in \RR^k$. In the classical situation\footnote{The symbols $\1_n$, $\0_n$, $\I_n$ and $\J_n$ denote the $n$-dimensional vector of ones, $n$-dimensional vector of zeros, the $n \times n$ unit matrix and the $n \times n$ matrix of ones, respectively.} $\A=\1_n^T$ and $\b=N \in \NN$; in that case we only restrict the number $N$ of trials, the so-called size of the experiment. Nevertheless, there are also many situations where $\A$ and $\b$ are more complex. They can correspond to various time, budget, material, unbiasedness and safety restrictions, or requirements on the form of the design (see, e.g., \citet{HBF} and Section \ref{sec:EX} of this paper). An important constraint necessary for applications to subsampling is the that each design point (i.e., data-point) can be used only once; formally $\A=\I_n$ and $\b=\1_n$.\footnote{In actual computation using integer programming solvers this ``without replication'' constraint can be forced by setting the type of variables to binary.} 
\bigskip

Suppose that the information gained from an experiment based on $\xi \in \Xi^E_{\A,\b}$ can be represented by a matrix $\M(\xi)$. For instance, $\M(\xi)$ may be proportional to the Fisher information matrix for the unknown parameters of an underlying statistical model. In optimal experimental design, it is usual to select a concave function $\Phi$ with a target set $\RR \cup \{-\infty\}$ to quantify the information content of $\M(\xi)$. Such an optimality criterion allows an experimenter to compare different designs and, in principle, to select a $\Phi$-optimal ED, i.e., a design that maximizes\footnote{Alternatively, it is possible to select a \emph{convex} criterion $\Phi$ such that $\Phi(\M(\xi))$ can be interpreted as a loss from the experiment that depends on the design $\xi$. In this case, the optimal design would minimize $\Phi(\M(\cdot))$ over $\Xi^E_{\A,\b}$. Note that some criteria do not depend on the design via its information matrix; we will not discuss them in this paper.} $\Phi(\M(\cdot))$ over the discrete set $\Xi^E_{\A,\b}$. In many cases, the information matrix can be consistently extended to ADs. Then, maximizing $\Phi(\M(\cdot))$ over the convex set $\Xi^A_{\A,\b}=\{ \xi \in \RR_+^n: \A\xi \leq \b\}$ results in the so-called optimal AD.
\bigskip

The construction of optimal EDs is typically a difficult problem of discrete optimization. There are two general approaches to computing optimal or nearly-optimal EDs (see, e.g., \citet{MWY} for a survey):
\begin{enumerate}[(i)]
  \item Convex computational methods, in which an optimal AD is first determined and a process called ``rounding'' is then used to obtain an ED;
  \item Computational methods of discrete optimization, including complete or partial enumeration methods, as well as various specialized or general-purpose solvers and heuristics of mathematical programming.
\end{enumerate} 

It is usually much simpler to determine an optimal AD than an optimal ED, both theoretically and computationally. Therefore, a large part of the literature is concerned only with approximate designs (cf., \citet{Puk}, \citet{Pazman}). Although ADs cannot be directly used for conducting experiments, relatively little attention has been paid to their conversion into efficient EDs. 
\bigskip

The standard methods for converting an AD into an ED are called rounding algorithms, developed for the classical, size-constrained problem. A rounding algorithm begins with an AD $\xi^*=(\xi_1^*,\ldots,\xi_n^*)^T \in \Xi^A_{\1_n^T,N}$ and extracts a vector $\mathcal{W}=(\xi^*_{i_1}, \ldots, \xi^*_{i_s})^T$ of positive weights, where $\{i_1,\ldots,i_s\}=\{i: \xi^*_i>0\}$ is the support of $\xi^*$ and $s$ is the size of the support. Then, typically using simple rules, the algorithm converts $\mathcal{W}$ into a vector $\left(\xi^\pm_{i_1},\ldots,\xi^\pm_{i_s}\right)^T \subset \NN_0^s$ such that $\sum_j \xi^\pm_{i_j}=N$ and, finally, transforms the vector of rounded weights into an ED that belongs to $\Xi^E_{\1_n^T,N}$.
\bigskip

The first notable rounding method was suggested by \citet{kiefer}, who formulated the rounding problem as the minimization of the maximum of the difference between the exact and approximate design weights. By using techniques similar to those applied in voting apportionment, \citet{PukRieder} arrived at a criterion-independent rounding algorithm known as efficient rounding (ER). More recent proposals include randomized rounding heuristics, e.g., proportional and pipage rounding, as well as incremental rounding, and bounds on the approximation ratios of the resulting designs have been presented (see \citet{Bouhtou} and \citet{Sagnol}). However, these methods are only applicable if the criterion function is submodular (e.g., $D$-optimality). We are not aware of rounding procedures for general $\Xi^E_{\A,\b}$, but for specific classes of constraints, it is not difficult to mimic the existing rounding procedures originally developed for the size-constrained problem. 
\bigskip

ER and its variants, although prevalent to this day, have several major drawbacks. First, for any positive coordinate of the initial AD, the value of the corresponding coordinate of the resulting ED is forced to be at least $1$. This implies the restriction $N \geq s$, which can completely prevent the application of ER if the support size of the AD is large. From the opposite perspective, if a design point is not present in the support of the AD, then ER cannot add a corresponding design point into the resulting ED. Moreover, ER does not account for any design criterion nor any underlying statistical model, although it is based on an optimal AD, which itself can strongly depend on the adopted criterion and model. In addition, for many statistical models, an infinite number of optimal ADs exist, and it is unclear which of them should be used for the rounding operation. All of these disadvantages generally make approach (ii) preferable to (i) in practice; see, e.g., the examples in \citet{GoosJones}. 
\bigskip

\citet{HF} proposed a substantially different approach to the use of an optimal AD for ED construction, which overcomes many disadvantages of ER and similar methods. In particular, it does not depend on the choice of the optimal AD if the AD is not unique, it is not restricted to the support of the optimal AD, and the resulting EDs are usually significantly more efficient than the EDs computed by ER. The method is based on a second-order approximation of the $D$-criterion in the neighborhood of the $D$-optimal approximate information matrix, and to arrive at an ED, it employs rapid off-the-shelf solvers for integer quadratic programming (IQP).
\bigskip

In this paper, we view the idea of a quadratic criterion approximation based on an optimal or nearly-optimal AD as a broadly applicable principle in computational experimental design, and we call this principle AQuA (ascent with quadratic assistance). As we will show, AQuA can be realized by means of heuristics but also via solvers of IQP or mixed integer conic quadratic programming (MICQP) in situations with various budget and structural constraints on the design. AQuA can also be used sequentially, similarly to the sequential quadratic programming.
\bigskip

The new results of this paper demonstrate that AQuA can be applied to a wide range of criteria, including the important criteria of $A$- and $I$-optimality, and, utilizing a low-rank property of key quadratic forms, to much larger design spaces than competing methods.
\bigskip

This paper is organized as follows: In Section \ref{sec:Kiefer}, we present the general statistical model that we consider and two versions of Kiefer's $\Phi_p$-criteria. Subsequently, in Section \ref{Sec:SQR}, we demonstrate how to compute quadratic approximations of these criteria. We propose a low-rank method for the efficient application of AQuA in Section \ref{sec:AQuA}. This leads to the main result of this paper, a MICQP formulation of AQuA that can be practically used for large structured or unstructured design spaces. Section \ref{Sec:misc} provides various remarks. Finally, Section \ref{sec:EX} presents examples of optimal designs that can be computed by the application of the AQuA approach.

\section{The model and Kiefer's criteria}\label{sec:Kiefer}
 
For a trial in $\x_i \in \X$, $i=1,\ldots,n$, the observed response $Y(\x_i)$ is an $r$-dimensional random vector that is assumed to satisfy the linear regression model $E(Y(\x_i))=\A^T_i \beta$, where $\beta \in \RR^m$ is a vector of unknown parameters and $\A_i \in \RR^{m \times r}$ is a known matrix. For different observations, the errors are assumed to be independent and identically distributed with a finite and non-zero variance. Note that we consider a linear regression model with homoscedastic errors only for simplicity. It is straightforward to use the results of this paper for the construction of locally optimal designs of non-linear regression models (it only requires a linearization in a nominal parameter of the model; see, e.g., \citet{Atkinson}, Chap. 17) and also to heteroscedastic observations (by means of a proper transformation of the model; see \citet{Atkinson}, Chap. 23).
\bigskip

The information matrix associated with a design $\xi$ on $\X$, either exact or approximate, is
\begin{equation*}
\M(\xi)=\sum_{i=1}^n\xi_i \H_i,
\end{equation*}
where the $\H_i=\A_i\A^T_i$, $i=1,\ldots,n$, are non-negative definite ``elementary'' information matrices with dimensions of $m\times m$\footnote{For brevity, we will henceforth use $\mathcal{S}^m$, $\mathcal{S}^m_+$, and $\mathcal{S}^m_{++}$  to denote the sets of all symmetric, non-negative definite and positive definite $m \times m$ matrices, respectively.}. For the classical case with univariate observations, $\A_i=\f(\x_i) \in \RR^m$, i.e., the $m$-dimensional regressor corresponding to $\x_i \in \X$. The general form of the elementary information matrices may be useful for instance for problems with grouped or multivariate observations with possibly correlated components (see \citet{Pazman}, Sec. II.5.3.), optimal augmentation of a set of existing trials (as shown in \citet{HT}, Section 6) and elsewhere.
\bigskip

Let $\Phi: \mathcal{S}^m_+ \to \RR \cup \{-\infty\}$ be a continuous optimality criterion that attains its smallest value for singular non-negative definite matrices. Note that a $\Phi$-optimal ED exists because $\Xi^E_{\A,\b}$ is finite and non-empty. Since $\Xi^A_{\A,\b}$ is a non-empty compact set, a $\Phi$-optimal AD is also guaranteed to exist. If $\xi^*$ is a $\Phi$-optimal exact (approximate) design, then $\M(\xi^*)$ is called a $\Phi$-optimal exact (approximate) information matrix. To make the optimal design problem non-trivial, we will suppose that there exists a $\xi \in \Xi^E_{\A,\b}$ such that $\M(\xi)$ is non-singular, which implies that both the approximate and exact $\Phi$-optimal information matrices are non-singular. We will also assume that $\Phi$ is twice differentiable in $\mathcal{S}^m_{++}$ and that there exists a version\footnote{By two versions of a criterion, we mean two criteria that induce the same ordering on the set of information matrices.} of $\Phi$ that is strictly concave in the optimal approximate information matrix $\M_*$. This assumption is satisfied for most models and standard optimality criteria, and it implies that the $\Phi$-optimal approximate information matrix is unique. 
\bigskip

All properties stated above are satisfied by Kiefer's criteria, which are commonly used in practice. In the optimal design literature, several versions of Kiefer's criteria for $\Phi_p$-optimality appear, and usually the choice of the particular version does not affect the strength of the theoretical or computational results. However, it turns out that in general, criterion-approximation methods \emph{do} depend on the particular version of the criterion that is chosen. Therefore, we will consider two concave versions of $\Phi_p$ criteria, as follows.
\bigskip

The ``positive'' version (cf. \citet{Puk}): For $p \in \NN$ and $\M \in \mathcal{S}^m_{++}$,
 let 
\begin{equation}\label{eq:Phiplus}
  \Phi^+_p(\M)=\left(\frac{1}{m}\tr (\M^{-p}) \right)^{-1/p}
\end{equation}
and $\Phi^+_p(\M)=0$ for a singular matrix $\M  \in \mathcal{S}^m_{+}$. In particular, for $p=1$, we obtain the criterion $\Phi^+_1$ of $A$-optimality. The corresponding criterion of $D$-optimality is defined as $\Phi^+_0(\M)=\left(\det(\M)\right)^{1/m}$ for all $\M  \in \mathcal{S}^m_{+}$.
\bigskip

The ``negative'' version (cf. \citet{Pazman}, Section IV.2.7): For $p \in \NN$ and $\M \in \mathcal{S}^m_{++}$, let  
\begin{equation}\label{eq:Phiminus}
\Phi^-_p(\M)=-\left(\frac{1}{m}\tr(\M^{-p}) \right)^{1/p}
\end{equation}
and $\Phi^-_p(\M)= -\infty$ for a singular matrix $\M  \in \mathcal{S}^m_{+}$. In particular, $\Phi^-_1$ is a version of the $A$ criterion, and the corresponding $D$ criterion is $\Phi^-_0(\M)=-\left(\det(\M)\right)^{-1/m}$ for $\M  \in \mathcal{S}^m_{++}$ or $\Phi^-_0(\M)= -\infty$ for a singular $\M  \in \mathcal{S}^m_{+}$. 
\bigskip

Another commonly used concave version of the $D$-optimality criterion is $\Phi^0_0(\M)=\log(\det(\M))$ for all $\M  \in \mathcal{S}^m_{+}$ ($\log(0):=-\infty$); cf. \citet{Pazman}. This is the version of $D$-optimality used in \citet{HF}.
\bigskip

Note that both positive and negative versions of the criterion are smooth on the set of positive definite matrices, and the gradients are
\begin{equation*}
\nabla_\M\Phi^\pm_p(\M)=\pm\frac{\Phi^\pm_p(\M)}{\tr (\M^{-p})} \M^{-p-1}.
\end{equation*}

It is customary to evaluate the quality of a design with respect to the optimal AD. Let $\xi^*$ be the optimal AD, and let $\Phi$ be a non-negative, positively homogeneous criterion that is not constantly equal to zero (these conditions are satisfied by the positive version $\Phi^+_p$ of Kiefer's criteria). Then, the $\Phi$-efficiency of a design $\xi$ is defined as $\mathrm{eff}_{\Phi}(\xi)=\frac{\Phi(\M(\xi))}{\Phi(\M(\xi^*))}$, see \cite{Puk}, Section 5.15.

\section{Quadratic approximations of Kiefer's criteria}\label{Sec:SQR}

Suppose that we have a quadratic approximation $\Phi_Q: \mathcal{S}^m_+ \to \RR$ of a concave criterion $\Phi$ in the neighborhood of $\M_*$. Our experience shows that in most optimal design problems, the ordering on $\Xi^E_{\A,\b}$ that is induced by $\Phi_Q$ largely coincides with the ordering induced by the original criterion $\Phi$. At the same time, the quadratic approximation criterion $\Phi_Q$ can be evaluated (or updated) much more rapidly than $\Phi$, it has a simpler analytic properties and there are powerful available solvers that can maximize $\Phi_Q$.
\bigskip

Let $\M_*\in\mathcal{S}^m_{++}$ denote the $\Phi$-optimal approximate information matrix\footnote{Note that the optimal approximate information matrix $\M_*$ with respect to $\Phi_p^+$ and $\Phi_p^-$ is non-singular for any $p \in \NN_0$.} and let $\Phi$ be twice differentiable in $\mathcal{S}^m_{++}$. Then, a second-order Taylor approximation of $\Phi$ in terms of $\M \in \mathcal{S}^m_{+}$ can be written as follows
(see, e.g., \citet{Dattorro}, Appendix D) 
\begin{eqnarray}
\Phi(\M) &\approx & \Phi(\M_*)+\partial\Phi(\M_*,\M-\M_*) \nonumber \\
         && +\frac{1}{2}\partial^2\Phi(\M_*,\M-\M_*), \label{taylor}
\end{eqnarray}
where $\partial\Phi(\M,\N)$ denotes the directional derivative at the point $\M$ in the direction $\N$, i.e., $\partial\Phi(\M,\N)=\tr\left(\nabla_\M\Phi(\M)\N\right)$, and $\partial^2\Phi(\M,\N)$ denotes the second directional derivative at the point $\M$ in the direction $\N$, i.e., $\partial^2\Phi(\M,\N)=\tr\left(\nabla_\M\partial\Phi(\M,\N)\N\right)$, with $\nabla_\M\Phi(\M)$ denoting the gradient with respect to $\M$. 
\bigskip

For $\Phi=\Phi_p^+$ as defined in \eqref{eq:Phiplus} with $p \in \NN_0$, we obtain
\begin{equation*}
\partial\Phi_p^+(\M_*,\M-\M_*)=\frac{\Phi_p^+(\M_*)}{\tr(\M_*^{-p})}\tr(\M_*^{-p-1}(\M-\M_*))
\end{equation*}
and
\begin{eqnarray*}
\partial^2\Phi_p^+(\M_*,\M-\M_*)=\frac{\Phi_p^+(\M_*)}{\tr(\M_*^{-p})}\Big[(p+1)\frac{\tr^2(\M_*^{-p-1}\M)}{\tr(\M_*^{-p})}\Big.\\ \Big. -\mathcal{F}_p(\M_*,\M, \M)\Big],
\end{eqnarray*}
where 
\begin{equation*}
\mathcal{F}_p(\M_*,\M_1,\M_2)=\sum\limits_{r=1}^{p+1}\tr(\M_*^{-r}\M_1\M_*^{-p-2+r}\M_2).
\end{equation*}
Note that in particular,
\begin{eqnarray*}
  \mathcal{F}_0(\M_*,\M, \M)&=&\tr([\M_*^{-1}\M]^2) \\
  \mathcal{F}_1(\M_*,\M, \M)&=&2\tr(\M_*^{-2}\M\M_*^{-1}\M).
\end{eqnarray*}
According to \eqref{taylor}, we have $\Phi_p^+(\M)\approx \Phi_{pQ}^+(\M)$, where $\Phi_{pQ}^+$ is the second-order approximation of the criterion of $\Phi_p^+$-optimality and is given by
\begin{eqnarray*}
    \Phi_{pQ}^+(\M)=\frac{\Phi_p^+(\M_*)}{\tr(\M_*^{-p})}\Big( \tr(\M_*^{-p-1}\M)+ \Big.\\
    \Big.\frac{p+1}{2}\frac{\tr^2(\M_*^{-p-1}\M)}{\tr(\M_*^{-p})}-\frac{1}{2}\mathcal{F}_p(\M_*,\M,\M)\Big).\nonumber
\end{eqnarray*}

Similar computations can be performed for $\Phi=\Phi_p^-$, $p \in \NN_0$, as defined in \eqref{eq:Phiminus}, leading to the quadratic approximation 
\begin{eqnarray*}
	\Phi_{pQ}^-(\M)=\frac{-3\Phi_p^-(\M_*)}{\tr(\M_*^{-p})}\Big(\tr(\M_*^{-p-1}\M)-\tr(\M_*^{-p})\Big.\\
	\Big. +\frac{p-1}{6}\frac{\tr^2(\M_*^{-p-1}\M)}{\tr(\M_*^{-p})}
	-\frac{1}{6}\mathcal{F}_p(\M_*,\M, \M)\Big).\nonumber
\end{eqnarray*}

We remark that it is also possible to compute $\Phi_{pQ}^-$ based on the formulas for the Hessian of a modified version of $\Phi_p^-$ that were derived by \citet{YBT} (for integer values of $p$) and \citet{SY} (for $p=0,1$).
\bigskip

Because the mapping $\xi \to \M(\xi)$ is linear, $\Phi_Q(\M(\cdot))$ is a quadratic function on $\RR^n_+$; i.e., $\Phi_Q(\M(\cdot))=\phi_{\h,\Q}(\cdot)+c$ for some $\h \in \RR^n$, $\Q \in \mathcal{S}^n_+$ and $c \in \RR$, where
\begin{equation}\label{eq:phihQ}
  \phi_{\h,\Q}(\xi)=\h^T\xi - \xi^T\Q\xi, \:\: \xi \in \RR^n_+.
\end{equation}

Then, the problem of optimal ED based on the AQuA approach can be expressed as the integer quadratic problem
\begin{equation}\label{opt}
\left.\begin{array}{rl}
\max_{\xi} & \phi_{\h,\Q}(\xi), \\
\hbox{subject to} & \xi\in\Xi^E_{\A,\b}.
\end{array}\right.
\end{equation}

For a general criterion, there are several possible ways of constructing the appropriate vector $\h$ and matrix $\Q$, for instance, through the use of standard numerical differentiation techniques. However, for Kiefer's criteria with $p \in \NN_0$, it is simple to derive analytical forms for $\h$ and $\Q$, as we show next.
\bigskip

Consider an ED $\xi=(\xi_1,\ldots,\xi_n)^T$ with the information matrix $\M=\M(\xi)$. Clearly, $\tr(\M_*^{-p-1}\M)=\sum_i \xi_i \tr(\M_*^{-p-1}\H_i)$, and for $r=1,\ldots, p+1$,
\begin{equation*}
 \mathcal{F}_p(\M_*,\M,\M)=\sum_{i,j=1}^n\xi_i \xi_j \mathcal{F}_p(\M_*,\H_i,\H_j);
\end{equation*}
therefore, the maximization of $\Phi_{pQ}^+(\M(\cdot))$ over $\Xi^E_{\A,\b}$ is equivalent to the integer quadratic optimization problem expressed in \eqref{opt}, where $\h=\h_p^+$ has the components $(\h_p^+)_i=\tr(\M_*^{-p-1}\H_i)$, $i=1,\ldots,n$, and the matrix $\Q=\Q_p^+$ has the elements\footnote{Note that the matrix $\Q_p^+$ is symmetric, as is the matrix $\Q_p^-$ defined below, because $\tr(\M_1\H_1\M_2\H_2)=\tr(\M_1\H_2\M_2\H_1)$ for the symmetric non-negative definite matrices $\M_1$, $\M_2$, $\H_1$, and $\H_2$.}
\begin{equation*}
(\Q_p^+)_{i,j}=\frac{p+1}{2}\frac{(\h_p^+)_i(\h_p^+)_j}{\tr(\M_*^{-p})}-\frac{1}{2}\mathcal{F}_p(\M_*,\H_i,\H_j),
\end{equation*}

$i,j=1,\ldots,n$. Similarly, the maximization of $\Phi_{pQ}^-(\M(\cdot))$ over $\Xi^E_{\A,\b}$ is equivalent to the integer quadratic optimization problem expressed in \eqref{opt}, where $\h=\h_p^-$ has the components $(\h_p^-)_i=\tr(\M_*^{-p-1}\H_i)$, $i=1,\ldots,n$, and the matrix $\Q=\Q_p^-$ has the elements
\begin{equation*}
(\Q_p^-)_{i,j}=\frac{p-1}{6}\frac{(\h_p^-)_i(\h_p^-)_j}{\tr(\M_*^{-p})}-\frac{1}{6}\mathcal{F}_p(\M_*,\H_i,\H_j),
\end{equation*}
$i,j=1,\ldots,n$.

\section{Efficient computational approach to AQuA}\label{sec:AQuA}

\subsection{A low-rank property of the quadratic approximations of Kiefer's criteria}\label{subsec:lowrank}

We can use quadratic approximation of criteria in combination with many algorithms for optimal ED (e.g., \citet{Atkinson}, \citet{Dykstra}, \citet{Haines}). To do so, we must be able to compute the values of the quadratic function $\phi_{\h,\Q}$ given in \eqref{eq:phihQ} for designs $\xi$, as required by the algorithm. We will show that for the quadratic approximation criteria resulting from the optimal ED problem based on the Kiefer's criteria, this computation can be performed rapidly, based on a low-rank property of the associated quadratic forms. As a key by-product, we will obtain a useful quadratic cone representation of the AQuA optimization problem.
\bigskip

The ability to efficiently numerically evaluate multivariate quadratic functions of the form $\phi_{\h,\Q}$ generally depends on various specifics of the problem at hand, the known theoretical properties of $\h$ and $\Q$, the selected optimization algorithm, and the available hardware. Here, we will consider problems that are typical of optimal experimental design. In particular, we will assume that $m$ is a small number (usually less than $10$), whereas $n$ is a much larger number, possibly ranging from the order of tens to hundreds of thousands.
\bigskip

Let the function $\phi_{\h,\Q}$ be based on the quadratic approximation of a criterion defined on the set of information matrices. That is, $\phi_{\h,\Q}(\cdot)=\Phi_Q(\M(\cdot))$, where $\Phi_Q(\M)$ is a quadratic function of the elements of $\M$. For a design $\xi$, the most problematic part of computing $\phi_{\h,\Q}(\xi)$ is the evaluation of the quadratic form $\xi^T\Q\xi$ for the $n \times n$ matrix $\Q$, because $n$ is often large. However, as we will show, we can construct a matrix $\SS$ with dimensions of $n \times t$, where $t \leq s:=m(m+1)/2 \ll n$, such that $\Q=\SS \SS^T$. Importantly, we can construct $\SS$ without computing $\Q$; i.e., we can completely avoid working with potentially enormous matrices.
\bigskip

To this end, let the function $\Phi_Q: \mathcal{S}^m_+ \to \RR$ be represented in the form   
\begin{equation*}
  \Phi_Q(\M)=a(\tilde{\h}^T\vech(\M) - (\vech(\M))^T\tilde{\Q}\:\vech(\M)) + c,
\end{equation*}
where $\tilde{\h} \in \RR^s$, $\tilde{\Q} \in \mathcal{S}^s_+$, and $a>0$, $c$ are real numbers which do not influence the maximum. Let $\G_m\in\mathbb{R}^{m^2\times s}$ be the duplication matrix that relates the $\vech$ and $\mathrm{vec}$ operators\footnote{The symbols $\vech$ and $\mathrm{vec}$ denote the vectorization and half-vectorization of a matrix, respectively.}; i.e., $\mathrm{vec}(\M)=\G_m \vech(\M)$. Then, the versions of Kiefer's criteria defined in the previous section can be represented using Theorem 16.2.2. from \citet{Harville} and the formulas
\begin{eqnarray*}
  \tr(\N\M)&=&(\mathrm{vec}(\N))^T\G_m \vech(\M), \\
  \tr^2(\N\M)&=&\vech(\M)^T \G_m^T \mathrm{vec}(\N)\\
             &&(\mathrm{vec}(\N))^T \G_m \vech(\M),\\
  \tr(\N_1\M\N_2\M)&=&\vech(\M)^T \G^T_m (\N_2\otimes \N_1)\\
             &&\G_m \vech(\M),
\end{eqnarray*}
which are valid for all $\N, \M, \N_1, \N_2 \in \mathcal{S}^m$; thus, we obtain
\begin{equation*}
  \Phi_{pQ}^{\pm}(\M)=a^{\pm}\left((\tilde{\h}_p^{\pm})^T\vech(\M) - \vech(\M)^T\tilde{\Q}_p^{\pm}\:\vech(\M)\right) + c^{\pm},
\end{equation*}
where 
\begin{equation*}
 a^+=\frac{\Phi_p^+(\M_*)}{\tr(\M_*^{-p})}, \:\: a^-=\frac{-3\Phi_p^-(\M_*)}{ \tr(\M_*^{-p})},
\end{equation*}
\begin{equation*}
 \tilde{\h}^+_p=\tilde{\h}^-_p=\G_m^T\mathrm{vec}(\M_*^{-p-1}),
\end{equation*}
\begin{eqnarray*}
  \tilde{\Q}^+_p&=&\G_m^T\Big[-\frac{1+p}{2}\frac{\mathrm{vec}(\M_*^{-p-1})(\mathrm{vec}(\M_*^{-p-1}))^T}{\tr(\M_*^{-p})}+\Big.\\
  &&\Big.\frac{1}{2}\sum_{r=1}^{p+1}\M_*^{-p-2+r}\otimes \M_*^{-r}\Big]\G_m,\\
  \tilde{\Q}^-_p&=&\G_m^T\Big[\frac{1-p}{6}\frac{\mathrm{vec}(\M_*^{-p-1})(\mathrm{vec}(\M_*^{-p-1}))^T}{\tr(\M_*^{-p})}+\Big.\\
  &&\Big.\frac{1}{6}\sum_{r=1}^{p+1}\M_*^{-p-2+r}\otimes \M_*^{-r}\Big]\G_m,
\end{eqnarray*}
and $c^+=0$, $c^-= 3\Phi_p^-(\M_*)$.
\bigskip

Next, we can construct a decomposition $\tilde{\Q}=\tilde{\C}\tilde{\C}^T$ such that the $s \times t$ matrix $\tilde{\C}$ is of rank $t$,\footnote{Note that $\tilde{\Q}$ can be a singular non-negative definite matrix; therefore, $t$ can be even smaller than $s$.} using, for instance, the Cholesky algorithm or the singular value decomposition. We have 
\begin{eqnarray*}
  \phi_{\h,\Q}(\xi)&=&\Phi_Q\left(\sum_{i=1}^n\xi_i\H_i\right)=\sum_{i=1}^n\xi_i \tilde{\h}^T\vech(\H_i) \nonumber \\
  &&- \sum_{i=1}^n\sum_{j=1}^n\xi_i\xi_j(\vech(\H_i))^T\tilde{\Q}\:\vech(\H_j)\nonumber\\
  &=&\xi^T (\H\tilde{\h})^T - \sum_{i=1}^n\sum_{j=1}^n\xi_i\xi_j(\H\tilde{\C}\tilde{\C}^T\H^T)_{i,j} \nonumber \\
  &=&\h^T\xi - \|\SS^T \xi\|^2,\label{eq:lrr}
\end{eqnarray*}
where $\H=(\vech(\H_1),\ldots,\vech(\H_n))^T$ is an $n \times s$ matrix, $\h=\H\tilde{\h}$, and $\SS=\H\tilde{\C}$. Equation \eqref{eq:lrr} allows us to compute $\phi_{\h,\Q}(\xi)$ without evaluating and storing $\Q$.
\bigskip

An advantage of the previous expression is that with the use of $\SS$, $\phi_{\h,\Q}(\xi)$ can be rapidly evaluated; for instance, the exchange step in an exchange algorithm (\citet{Atkinson}, Sec. 12.3) can be performed based on the equation
\begin{eqnarray*}
 &&\phi_{\h,\Q}(\xi+\e_l-\e_k)=\\
 &&\phi_{\h,\Q}(\xi)+\h_l-\h_k-2(\SS^T\xi)^T[\SS_{l\cdot}- \SS_{k\cdot}]-\\ 
 &&\|\SS_{l\cdot}\|^2+2(\SS_{l\cdot})^T\SS_{k\cdot}-\|\SS_{k\cdot}\|^2,
\end{eqnarray*}
where $\e_l$, $\e_k$ are the $l$-th and $k$-th standard unit vectors, and $\SS_{l\cdot}$, $\SS_{k\cdot}$ are the $l$th and the $k$th rows of $\SS$. Note that $\SS^T\xi$ is updated as follows: $\SS^T(\xi+\e_l-\e_k)=\SS^T\xi+\SS_{l\cdot}-\SS_{k\cdot}$. If the $n$ values of $\|\SS_{k\cdot}\|^2$, $k=1,\ldots,n$, are precomputed and stored in memory, then each update involves only $2t+2$ multiplications and $4t+4$ subtractions or additions. An example of how these formulas can be utilized with heuristic exchange algorithm can be found in a preprint of the previous version of this paper; see \citet{aqua}. Here we will focus on a more versatile application of the low-rank property, as detailed in the next section.

\subsection{Mixed integer conic quadratic programming formulation of AQuA}\label{Sec:MICQP}

Once we proved the low-rank property $\Q=\SS\SS^T$, where $\SS$ is an $n \times t$ matrix, $t \ll n$, we can use a known trick to reformulate the problem of quadratic programming (e.g., \cite{moseka}, Chapter 10). Introducing an auxiliary continuous variable $r$, the optimization problem \ref{opt} of the AQuA approach can be written as
\begin{equation}\label{prob:aux}
	\left.\begin{array}{rl}
	\max_{\xi, r} & \h^T\xi - r\\
	\hbox{s.t.} & \A\xi\leq\b,\ \xi\geq \0_n,\ \xi \in \mathbb{Z}^n\\
	  & r \geq ||\SS^T\xi||^2.
	\end{array}\right.
\end{equation}
	
It is simple to verify that the last constraint in \eqref{prob:aux} can be expressed as 
$\V(1/2,r,\xi^T\SS)^T \in Q^{2+t}$, where $Q^{2+t}$ is the second-order cone
\begin{equation*}
 Q^{2+t}=\{(a,b,\v^T)^T: a \geq \|(b,\v^T)^T\| \}
\end{equation*}
and $\V$ is the orthogonal matrix
\begin{equation*}
       \V=
  \left( {\begin{array}{ccc}
   \frac{1}{\sqrt{2}} & \frac{1}{\sqrt{2}} & \0_t^T \\
   \frac{1}{\sqrt{2}} & -\frac{1}{\sqrt{2}} & \0_t^T \\
   \0_t & \0_t & \I_t
  \end{array} } \right).
\end{equation*}
	
We thus obtained a mixed integer conic quadratic problem (MICQP) which can be formulated as follows: 
	\begin{equation}\label{aqua.cone}
	\left.\begin{array}{rl}
	\max_{\xi,\v,r,a,b} & \h^T\xi - r\\
	\hbox{s.t.} & \A\xi\leq\b,\ \xi\geq \0_n,\ \xi \in \mathbb{Z}^n\\
	  & 2\sqrt{2}a-2r =1, \\
	  & 2\sqrt{2}b+2r =1, \\
	  & \SS^T\xi-\v=\0_t, \\
	  & (a,b,\v^T)^T \in Q^{2+t}.
	\end{array}\right. 
	\end{equation}
	
  Note that the formulation \eqref{aqua.cone} has a linear objective function and does not require potentially huge $n \times n$ matrix $\Q$ at all; it only requires the $n \times t$ matrix $\SS$ which is often much smaller in optimum design problems. Moreover, the number of variables of \eqref{aqua.cone}, $n+t+3$, is only marginally larger than the number $n$ of variables in the direct integer quadratic formulation \eqref{opt}. Indeed, for $t \ll n$ our numerical studies prove that the formulation \eqref{aqua.cone} can be significantly more computationally efficient than \eqref{opt}, as we will demonstrate in Section \ref{sec:EX}.

\section{Miscellaneous comments}\label{Sec:misc}

\subsection{Continuous design spaces}\label{Subs:cont}

In some applications, it is possible to use a continuous design space $\tilde{\X}$, instead of a finite one. This is typical of factor experiments under the theoretical assumption that the levels of some factor can be any real numbers in a given interval. In such cases, AQuA cannot be directly applied\footnote{Of course, the same is true for a multitude of other popular design algorithms which work only on finite spaces.}.
However, a straightforward strategy is to first apply AQuA to a finite subset of $\tilde{\X}$, and then use its result as an initial design for any constrained continuous optimization method which adjusts the positions of the support points within $\tilde{\X}$. Note that the search for optimal positions of design points in a continuous space is generally a highly non-convex problem, and a good initial feasible solution provided by a finite-space method such as AQuA can make a crucial difference. See Subsection 5.1 in \cite{aqua} that demonstrates this approach for the full quadratic model ($m=15$) with $4$ continuous factors, i.e., $\tilde{\X}=[-1,1]^4$.

\subsection{Quadratic approximations of different versions of the same criterion}\label{Sec: gamma}

We can regard criteria $\Phi^+_p$ and $\Phi^-_p$ as part of a larger class of concave criteria: for $\gamma \in [-1,1]$ and for $p \in \NN_0$ we can define 
\[
\Phi^{(\gamma)}_p:=(1+\gamma)\Phi^+_p/2+(1-\gamma)\Phi^-_p/2,
\] where we set $0 \times \infty=0$.
Thus, $\Phi^+_p=\Phi^{(+1)}_p$ and $\Phi^-_p=\Phi^{(-1)}_p$ for all $p \in \NN_0$. Clearly, $\Phi^{(\gamma)}_p$ is a concave version of the same criterion for all $\gamma \in [-1,1]$ and its quadratic approximation is 
\[
\Phi^{(\gamma)}_{pQ}=(1+\gamma)\Phi^+_{pQ}/2+(1-\gamma)\Phi^-_{pQ}/2. 
\]
Note that setting $p=0$ and $\gamma=\gamma_d=\frac{1-d^2}{1+d^2}$, where $d=(\det \M_*)^{1/m}$, leads to the optimization problem of the form \eqref{opt} with
\begin{equation*}
\tilde{\h}_0=\G_m^T\mathrm{vec}(\M_*^{-1})
\end{equation*}
and
\begin{equation*}
\tilde{\Q}_0=\frac{1}{4}\G_m^T\left[\M_*^{-1}\otimes \M_*^{-1} \right]\G_m.
\end{equation*}
It is straightforward to verify that this choice of $\tilde{\h}_0$ and $\tilde{\Q}_0$ corresponds to the same quadratic approximation as the one that can be obtained from the $D$-optimality criterion in the form $\log(\det(\M))$, used in \citet{HF}. Note that we always have $\gamma_d \in (-1,1)$. That is, in the sense of the AQuA approach, the $\log\det$ criterion is always ``between'' the positive and the negative versions of $D$-optimality. 
\bigskip

Different versions of the same criterion lead to different quadratic approximations. Nonetheless, our numerical observations suggest that the differences are minor (see Subsection \ref{SBW}).

\subsection{Generalization of $I$-optimality and its conversion into $A$-optimality} \label{Sec:IV}
  
Recently, there has been much interest in $I$-optimality\footnote{This criterion is sometimes called called $IV$- or $V$-optimality (see Section 10.6 in \citet{Atkinson}).}, because $I$-optimality may be a more appropriate criterion than $D$-optimality if we are interested in the estimation of the mean value of the response (see, e.g., \citet{Mont}, \citet{LN}, and \citet{ABM}).
  
The results for $A$-optimality can be easily adapted to compute $I$-optimal designs. Standard $I$-optimal designs are applied to models with one-dimensional observations ($r=1$), and they minimize the integral of the variances of the BLUEs of the response surface over a region $\mathfrak{Y}$ with respect to some measure. We will generalize the notion of $I$-optimal design to potentially multivariate observations and show that $I$-optimal designs are $A$-optimal in a transformed model, giving us the possibility to use the theory and algorithms developed for $A$-optimality. 
\bigskip

 Let $\mathfrak{Y} \subseteq \mathbb{R}^d$ be a measurable set representing a region of prediction interest, and let $\eta$ be a measure on $\mathfrak{Y}$. Suppose that for each $\x \in \mathfrak{Y}$, there is a matrix $\V(\x) \in \mathcal{S}^m_+$ such that $\tr(\M^{-1}\V(\x))$ is a measure of variance of the response surface estimator in $\x$, provided that the information matrix for the parameters is $\M \in \mathcal{S}^m_{++}$. For a positive definite $\M$, we can define a (generalized) $I$-optimality criterion
\begin{equation*}
  \Phi_{I}(\M)=-\int_{\x \in \mathfrak{Y}}\tr(\M^{-1}\V(\x))\mathrm{d}\eta(\x)
  =-\tr\left(\M^{-1}\L\right),
\end{equation*}
where $\L=\int_{\x \in \mathfrak{Y}}\V(\x)\mathrm{d}\eta(\x)$, and for a singular $\M$, we can set $\Phi_{I}(\M)=-\infty$. Suppose that $\L=\mathbf{S}\mathbf{S}^T$, where $\mathbf{S}$ is non-singular. Then, clearly, a design $\xi$ is $I$-optimal if and only if it is $A$-optimal in the model given by the elementary information matrices 
\begin{equation*}
 \mathbf{S}^{-1}\H_1(\mathbf{S}^T)^{-1},\ldots,\mathbf{S}^{-1}\H_n(\mathbf{S}^T)^{-1}.
\end{equation*} 
 The standard situation corresponds to $r=1$, $\mathfrak{Y}=\X$, $\V(\x)=\f(\x)\f^T(\x)$, and $\eta$ being a uniform measure on $\X$.
\bigskip

We demonstrate the computation of $I$-optimal designs using AQuA in Subsections \ref{ss:scheffe} and \ref{ss:wine}.

\subsection{Iterative application of AQuA}\label{iter}

The central idea of this paper is to apply integer quadratic programming to a problem constructed on the basis of the optimal approximate information matrix $\M_*$, which is often available, either theoretically or via an efficient algorithm of convex optimization. Note, however, that the approximation is quite precise even if the criterion is based on a matrix $\tilde{\M}_*$ (henceforth called the 'anchor matrix') which is not perfectly optimal. Thus, in more difficult situations, in particular with a large design space and complex design restrictions, when $\M_*$ may be difficult to compute, we suggest to apply the following heuristic iterative scheme, similar to the successive application of the Newton's method to sequential quadratic optimization: 
\begin{enumerate}
  \item Compute a rough estimate $\tilde{\M}^{(0)}_*$ of $\M_*$ at a random subsample of $\X$ or neglecting some design constraints. Set $j$ to $0$.
  \item Use AQuA with the anchor matrix $\tilde{\M}^{(j)}_*$ instead of $\M_*$.\footnote{If this is not last iteration of the algorithm, we can use AQuA without the integer constraints on the design. Indeed this iterative approach can also be used for computing optimal \emph{approximate} designs, but we do not explore this possibility here.} Set $\tilde{\M}^{(j+1)}_*$ to be the information matrix of the resulting design.
  \item If a stopping rule is not satisfied, increase $j$ by one, and continue with the previous step.
\end{enumerate}

The previous scheme uses a sequence of successive quadratic optimization problems, which, in some cases, can be solved via the conic formulation of AQuA, despite the fact that we cannot solve the original optimal approximate problem because of its size or complexity. In the last subsection of the next section we will demonstrate that this approach can indeed lead to efficient EDs for large design spaces.  

\subsection{Current limitations of AQuA}

AQuA can be a valuable tool in the toolbox of computational methods of experimental design as numerically demonstrated in Section \ref{sec:EX}. However, it has currently no theoretical underpinnings in the sense of lower bounds on the efficiency of the resulting designs depending on general properties of the problem at hand\footnote{Note that after we already have a candidate exact design for a specific problem, we can compute a lower bound on its efficiency relative to the optimal approximate design. This often leads to a guarantee which is fully satisfactory for practical purposes. Moreover, many optimization heuristics which are eminently useful across sciences also lack theoretical bounds on the efficiency of the results that they generate.} Note that we have observed that AQuA sometimes produces significantly suboptimal designs for small design sizes $N \geq m$, in particular for $N=m$\footnote{See \citet{HF} for an example a strongly suboptimal result of AQuA for $N=m$.}. Moreover, we do not have a theoretical proof of convergence of the sequential approach outlined in Subsection \ref{iter}.
\bigskip

With easily available hardware and software, the IQP formulation of AQuA can solve problems with middle size $n$ (up to thousands) and any $m \leq n$. On the other hand, the MICQP version of AQuA can solve ``tall'' problems with a large $n$ (up to hundreds of thousands) and a relatively small $m \ll n$. However, we do not know how to use AQuA to handle problems with both $n$ and $m$ large.

\section{Numerical studies}\label{sec:EX}

The principle of AQuA can be applied to a wide spectrum of optimal design problems in various creative ways. Here we will choose several very different examples to inform the reader about general properties of AQuA, for instance:
\begin{enumerate}
\item the degree of reliability in achieving the optimal ED and the robustness with respect to the anchor matrix;
\item the possibility to efficiently construct solutions to optimal ED problems with complex constraints on the structure of the design;
\item the possibility to apply the conic version of AQuA to specific problems with a large design space, in particular to the problem of an information-based sub-selection of ``tall'' datasets.
\end{enumerate}

We will demonstrate the application and explore the performance of AQuA in the \texttt{R} computing environment (\citet{R}) employing the packages \texttt{OptimalDesign} (\citet{RLIB}), \texttt{matrixcalc} (\citet{matrixcalc}), and the mathematical programming solvers of \texttt{gurobi} (\citet{gurobi}). Note that there are also several other professional solvers that can handle IQP and MIQCP problems, for instance \texttt{mosek} (\citet{mosekb}). The examples were computed on a 64-bit Windows 10 system with an Intel Core i5-5500U processor at 2.40 GHz and 8 GB of RAM. The codes and additional information can be found at

\noindent \hyperlink{http://www.iam.fmph.uniba.sk/ospm/Harman/design/}{http://www.iam.fmph.uniba.sk/ospm/Harman/design/}.

For the application of the provided R codes, the user only needs to create the model (the matrix of all possible regressors $\f(x)$), the constraints (in the form of $\A$ and $\b$), and choose the criterion ($D$, $A$ or $I$).

\subsection{Size-constrained $D$- and $A$-optimal exact designs for the model of spring balance weighing of $6$ items}\label{SBW}

Consider the linear regression of the first degree without an intercept term on the vertices of the $m$-dimensional unit cube given by the formula
\begin{equation}\label{sbw}
E(Y(\x))=x_{1}\beta _1+\ldots+x_{m}\beta_m,
\end{equation}
where the components $x_{j}$ of $\x$ are chosen to be either $0$ or $1$. In \eqref{sbw}, the measurement $Y(\x)$ can be interpreted as the result of the weighing of items with unknown weights $\beta_1,\ldots,\beta_m$ on a spring balance, where $x_{j}$ denotes the presence or the absence of the item $j$. Here, the design space is the set of $n=2^m$ vertices of the unit cube in $\RR^m$. For this example, we selected $m=6$ items, that is, $n=64$.
\bigskip
 
The AD theory for model \eqref{sbw} with the standard constraint on the size of the experiment has been worked out in great detail: see, e.g., \citet{Cheng}, who used the equivalence theorem to find $\Phi_p$-optimal ADs for all values of $p$. For the application of AQuA, we can use the well-known ``neighbor vertex'' $D$-optimal and $A$-optimal ADs as described in \citet{Puk}, Sec. 14.10. For non-normalized ADs of size $N$, and for $s\in[0,m]$, the neighbor vertex design is
\begin{equation*}
  \xi_s=(1-(s-\lfloor s \rfloor))\zeta_{\lfloor s \rfloor} + (s-\lfloor s \rfloor)\zeta_{\lfloor s \rfloor+1},
\end{equation*}
where $\zeta_j$ is a $j$-vertex design, i.e., $\zeta_j$ assigns $N / \binom{m}{j}$ to the vertices of $\X$ having $j$ components equal to $1$ and $m-j$ components equal to $0$ and $\lfloor s \rfloor$ denotes the largest integer not exceeding $s$. For our case of $m=6$, the design $\xi_s$ with $s=\frac{24}{7}$ is $D$-optimal, its support size is $35$ and its information matrix is $\M^*_D=\frac{2N}{7} \I_6+\frac{2N}{7}\J_6$. Similarly, the design $\xi_s$ with $s=3$ is $A$-optimal, its support size is $20$ and its information matrix is $\M^*_A=\frac{3N}{10}\I_6+\frac{2N}{10}\J_6$.
\bigskip

In this model, the optimal ADs are not unique; the designs from Tables \ref{T:SBWD7} and \ref{T:SBWA7} are evidently not neighbor vertex designs, yet they are $D$- and $A$-optimal, respectively, which can be directly verified. Notice that the $D$-optimal approximate design from Table \ref{T:SBWD7} is evidently a $D$-optimal exact of size $N=7k$, $k \in \mathbb{N}$, and the $A$-optimal approximate design from Table \ref{T:SBWA7} is evidently an $A$-optimal exact design of size $N=10k$, $k \in \mathbb{N}$.\footnote{We stress that it is not completely trivial to find these balanced small-support $D$-, and $A$-optimal ADs in class of all optimal ADs; in fact, we have found them using the integer programming capabilities of AQuA. In this respect, AQuA can be very useful also for the problem with a single size constraint.}
\bigskip

We remark that, according to our experience, for a problem of optimal ED constrained only by the experimental size, well implemented heuristics such as the KL-exchange algorithm (\cite{Atkinson}, Section 12.6) will often outperform methods based on IP solvers, including AQuA, in terms of time required to achieve a practically optimal design. However, the existing heuristics and theoretical results for the selected size-constrained problem provide benchmarks that can be used to assess the properties of the AQuA method, as we show next.
\bigskip

For the numerical study of ED, we will use the experimental sizes of $N=6,7,\ldots,30$. For $m=6$, the $D$-optimal EDs are theoretically known (see \citet{NWZ}). For $A$-optimality and $m=6$ items, we are not aware of any publication which provides optimal EDs; therefore, we have computed the $A$-optimal EDs using the KL heuristic. We tested the AQuA approach realized by the integer quadratic solver of gurobi against the exact optimal values. To anchor the quadratic approximations, we used either the theoretically known optimal approximate information matrix $\M_*$, or a perturbed information matrix $\tilde{\M}_*$ that corresponds to a random design with efficiency $0.95$. The sub-optimal anchor matrix allows us to assess the robustness of the AQuA approach for problems where precise optimal AD is unavailable.  
\bigskip

\begin{figure}[!h]
\begin{center}
\includegraphics[width=\linewidth]{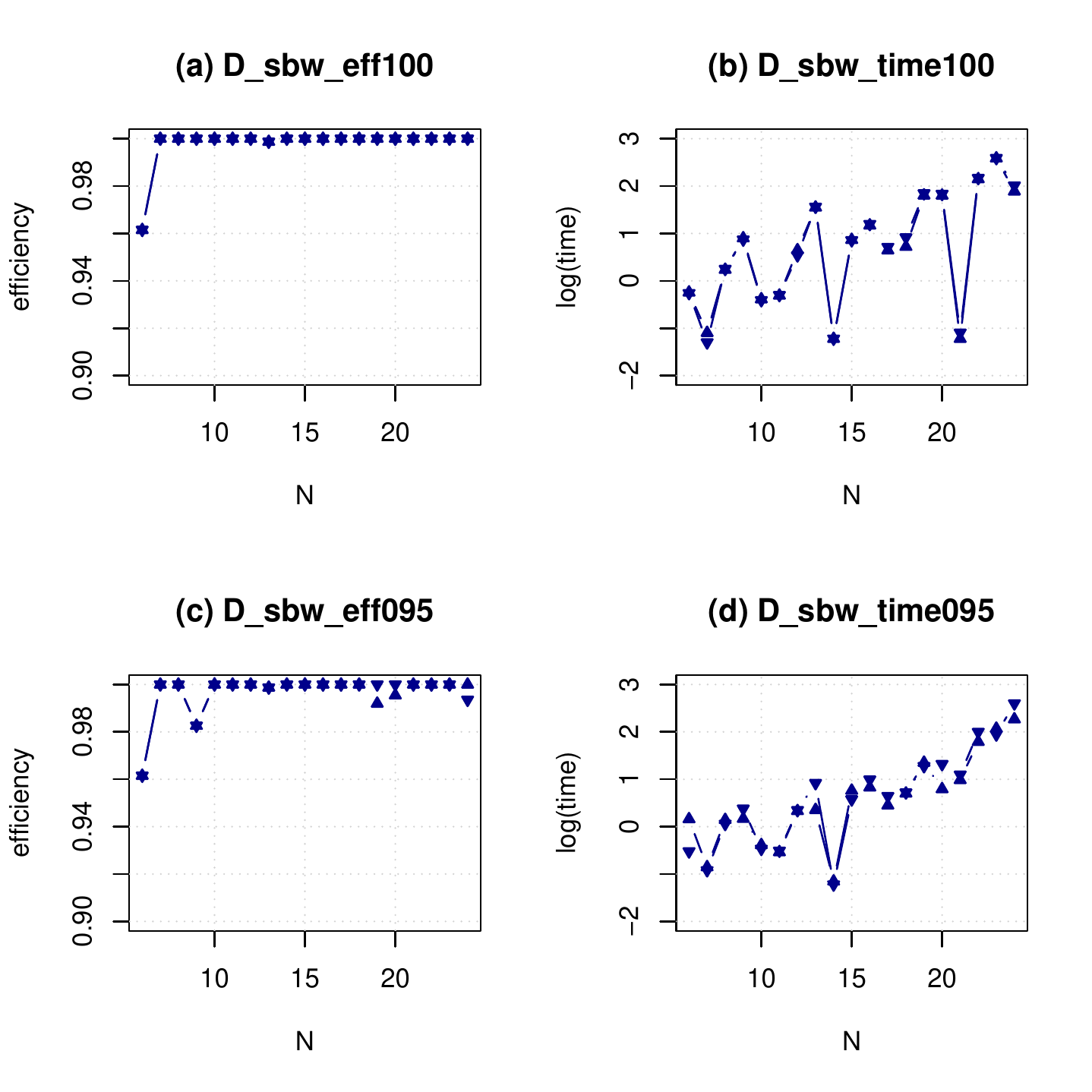}
	\caption{The efficiency and the computation times (in decadic logarithmic scale) of $D$-efficient designs for the model \eqref{sbw} with $m=6$ items and various numbers $N$ of measurements, as obtained via the direct IQP formulation of AQuA. The positive version of the quadratic approximation is denoted by $\bigtriangleup$ and the negative version of the quadratic approximation is denoted by $\bigtriangledown$. See the main text for details and discussion.}\label{F:sbwD}
\end{center}
\end{figure}

\begin{figure}[!h]
\begin{center}
\includegraphics[width=\linewidth]{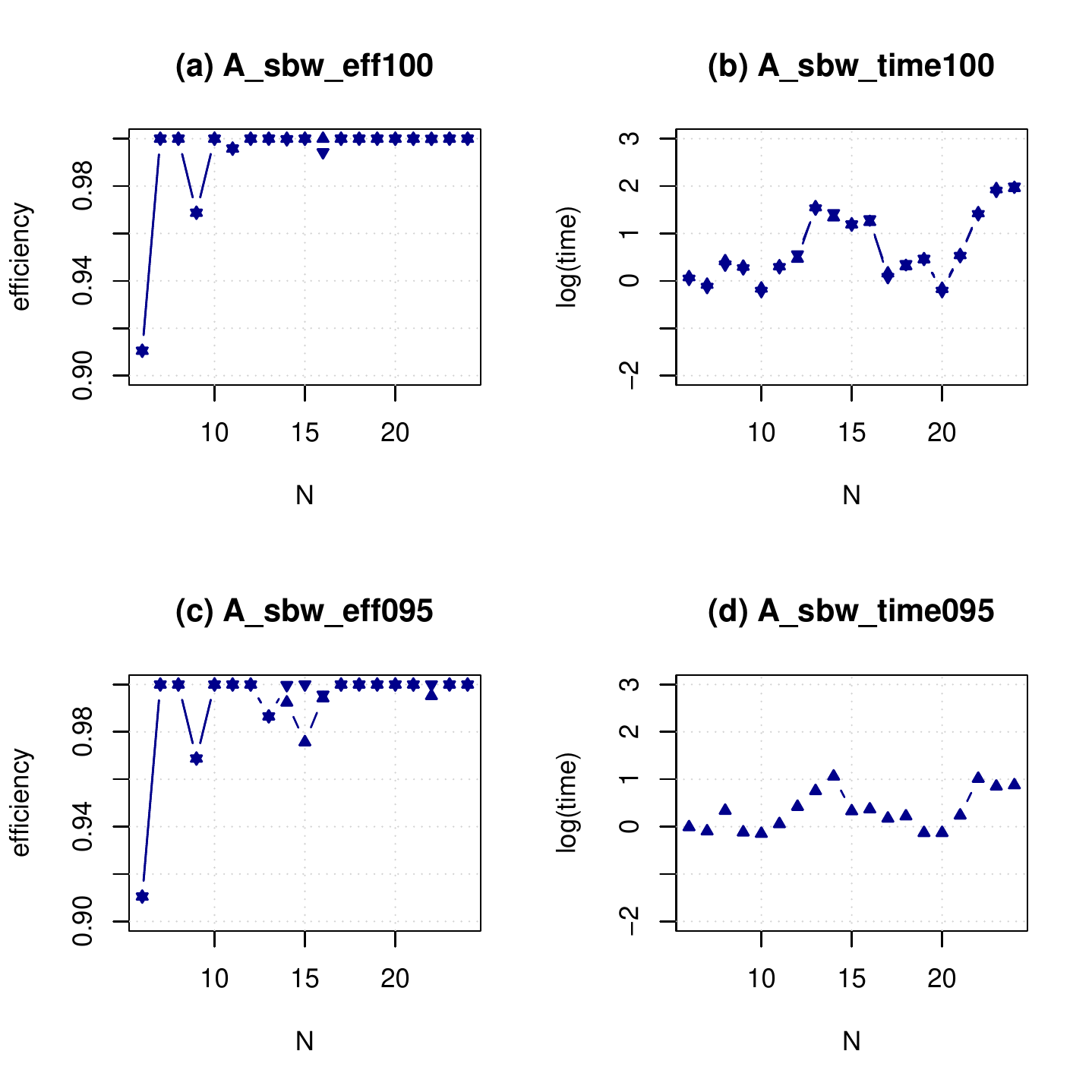}
	\caption{The efficiency and the computation times (in decadic logarithmic scale) of $A$-efficient designs for the model \eqref{sbw} with $m=6$ items and various numbers $N$ of measurements, as obtained via the direct IQP formulation of AQuA.  The positive version of the quadratic approximation is denoted by $\bigtriangleup$ and the negative version of the quadratic approximation is denoted by $\bigtriangledown$. See the main text for details and discussion.}\label{F:sbwA}
\end{center}
\end{figure}

The results, visualized in Figures \ref{F:sbwD} and \ref{F:sbwA}, can be summarized as follows:
\begin{itemize}
\item If $\M_*$ is precise (see the top panels of Figs. \ref{F:sbwD}, \ref{F:sbwA}), AQuA usually provides not only good, but perfectly optimal EDs. Less efficient results tend to occur for smaller sizes of $N$, in particular for $N=m$. 
\item The time to compute the solution generally increases with $N$ (see the right panels of Figs. \ref{F:sbwD}, \ref{F:sbwA}). However, if there is an optimal AD that coincides with an optimal ED of a given size, the computation tends to be rapid, in particular if $\M_*$ is precisely computed.
\item AQuA is generally robust with respect to the choice of the anchor matrix (see the bottom panels of Figs. \ref{F:sbwD}, \ref{F:sbwA}). Even using a significantly sub-optimal anchor matrix $\tilde{\M}_*$, the resulting EDs are either perfectly optimal or reasonably efficient, without a significant increase of the computation time (except a few specific cases of $N$ as discussed in the comment above).
\item There are some numerical differences between the two approximations of the $D$- and $A$-criteria, but they do not tend to be pronounced.
\end{itemize}

Note that the reported computation time corresponds to the moment at which the solver determines that its current design is good enough with respect to the quadratic criterion\footnote{We did not alter the default stopping rules and other options of the gurobi solver.}; the actual time that the solver first obtains the resulting design may be shorter.
\bigskip

It is also worth noting that the standard ER procedure cannot be applied to the neighbor vertex optimal ADs, for $N < 35$ in case of $D$-optimality, and for $N<20$ in case of $A$-optimality. The reason is that the neighbour vertex ADs have too many support points for ER to be applicable. Even in the remaining cases that ER can be applied, for instance if we used some auxiliary tools to obtain optimal ADs with a smaller support (such as those in Tables \ref{T:SBWD7} and \ref{T:SBWA7}), our computational experience suggests that the resulting EDs tends to be worse than those found by AQuA.

\begin{table}[!h]
\begin{center}
  \begin{tabular}{cccccc|c}
  $x_1$ & $x_2$ & $x_3$ & $x_4$ & $x_5$ & $x_6$ & $\xi^*_D(\x)$\\
  \hline
  1 & 1 & 0 & 1 & 0 & 0 & $N/7$ \\
  0 & 0 & 1 & 1 & 1 & 0 & $N/7$ \\
  0 & 1 & 1 & 0 & 0 & 1 & $N/7$ \\
  1 & 0 & 0 & 0 & 1 & 1 & $N/7$ \\
  1 & 1 & 1 & 0 & 1 & 0 & $N/7$ \\
  1 & 0 & 1 & 1 & 0 & 1 & $N/7$ \\
  0 & 1 & 0 & 1 & 1 & 1 & $N/7$
  \end{tabular}
  \caption{A $D$-optimal AD of size $N$ for the model from Subsection \ref{SBW}}\label{T:SBWD7}
\end{center}
\end{table}

\begin{table}[!h]
\begin{center}
  \begin{tabular}{cccccc|c}
  $x_1$ & $x_2$ & $x_3$ & $x_4$ & $x_5$ & $x_6$ & $\xi^*_A(\x)$\\
  \hline
  1 & 1 & 0 & 1 & 0 & 0 & $N/10$\\
  1 & 0 & 1 & 1 & 0 & 0 & $N/10$\\
  1 & 0 & 1 & 0 & 1 & 0 & $N/10$\\
  0 & 1 & 1 & 0 & 1 & 0 & $N/10$\\
  0 & 1 & 0 & 1 & 1 & 0 & $N/10$\\
  1 & 1 & 0 & 0 & 0 & 1 & $N/10$\\
  0 & 1 & 1 & 0 & 0 & 1 & $N/10$\\
  0 & 0 & 1 & 1 & 0 & 1 & $N/10$\\
  1 & 0 & 0 & 0 & 1 & 1 & $N/10$\\
  0 & 0 & 0 & 1 & 1 & 1 & $N/10$
  \end{tabular}
  \caption{An $A$-optimal AD of size $N$ for the model from Subsection \ref{SBW}}\label{T:SBWA7}
\end{center}
\end{table}

\subsection{Marginally constrained symmetric $D$- and $I$-optimal exact designs for the $3$-component Scheff\'{e} mixture model}\label{ss:scheffe}

The most important applications of AQuA can be expected in those situations for which there are no specialized heuristics, such as for problems with complex constraints on the design weights\footnote{We would like to stress that here we do not focus on the constraints on the design region, which are trivial to incorporate (at least in the case of finite design spaces); we work with constraints on the design vector itself in the polyhedral set of designs in $\RR^n$.}. For models with general linear constraints, few options are available if one wishes to find informative EDs. Namely, \citet{SagnolHarman} have shown that the $D$- and $A$-optimal EDs under general linear constraints on the design weights can be obtained by solving a specific mixed integer second-order cone programming problem (MISOCP). This approach, although, given enough time, it is guaranteed to find a perfectly optimal ED, is practically feasible only for problems that are small to medium in size (with currently common hardware, up to a thousand design points even with $m<10$). In this subsection, we will demonstrate that our approach can be superior to both the method of \citet{SagnolHarman}, as well as the direct application of a quadratic approximation as suggested in \citet{HF}.
\bigskip

Consider a mixture of three components with the ratio of each varying between 0\% and 100\% in increments of 2.5\%. The response can be modeled by a quadratic Scheff\'{e} mixture model given by
\begin{equation}\label{scheffe}
E(Y(\x))=\sum_{j=1}^3 \beta_j x_j+\sum_{u<v} \beta_{(uv)}x_u x_v, 
\end{equation}
where $\x=(x_1,x_2,x_3)$, $x_j\in\{0,0.025,0.05,\ldots,1\}$, $j=1,2,3$. Hence, the model contains $m=6$ unknown parameters, and the dimensionality of the set of designs $n=861$ (for more details and applications of mixture designs see, e.g., \citet{Cornell} and \citet{GJS}).
\bigskip

Suppose that, in addition to the size constraint, we are required to compute a design that fulfils a set of marginal constraints which require that each level of each factor can be used at most once, i.e., for all permissible designs $\xi$ and all $\tilde{\x}=(\tilde{x}_1,\tilde{x}_2,\tilde{x}_3) \in \X$ we have $\sum_{x_2,x_3} \xi(\tilde{x}_1,x_2,x_3) \leq 1$, $\sum_{x_1,x_3} \xi(x_1,\tilde{x}_2,x_3) \leq 1$ and $\sum_{x_1,x_2} \xi(x_1,x_2,\tilde{x}_3) \leq 1$. These ``non-collapsibility'' constraints can be justified similarly as the Latin hypercube designs (\citet{LHD}) and ``bridge'' designs (\citet{bridge}) on cubes. In particular, they lead to designs without replications of design points, which is important for computer experiments. Additionally, we have imposed  constraints of the form $\xi(x_1,x_2,x_3)=\xi(x_2,x_3,x_1)=\xi(x_3,x_1,x_2)$ that force the design to be symmetric. Therefore, we aim to find an optimal exact  design which combines properties of non-collapsibility of individual factor levels, symmetry, and efficiency of parameter estimation. 
\bigskip

\begin{figure}[!h]
\begin{center}
\includegraphics[width=\linewidth]{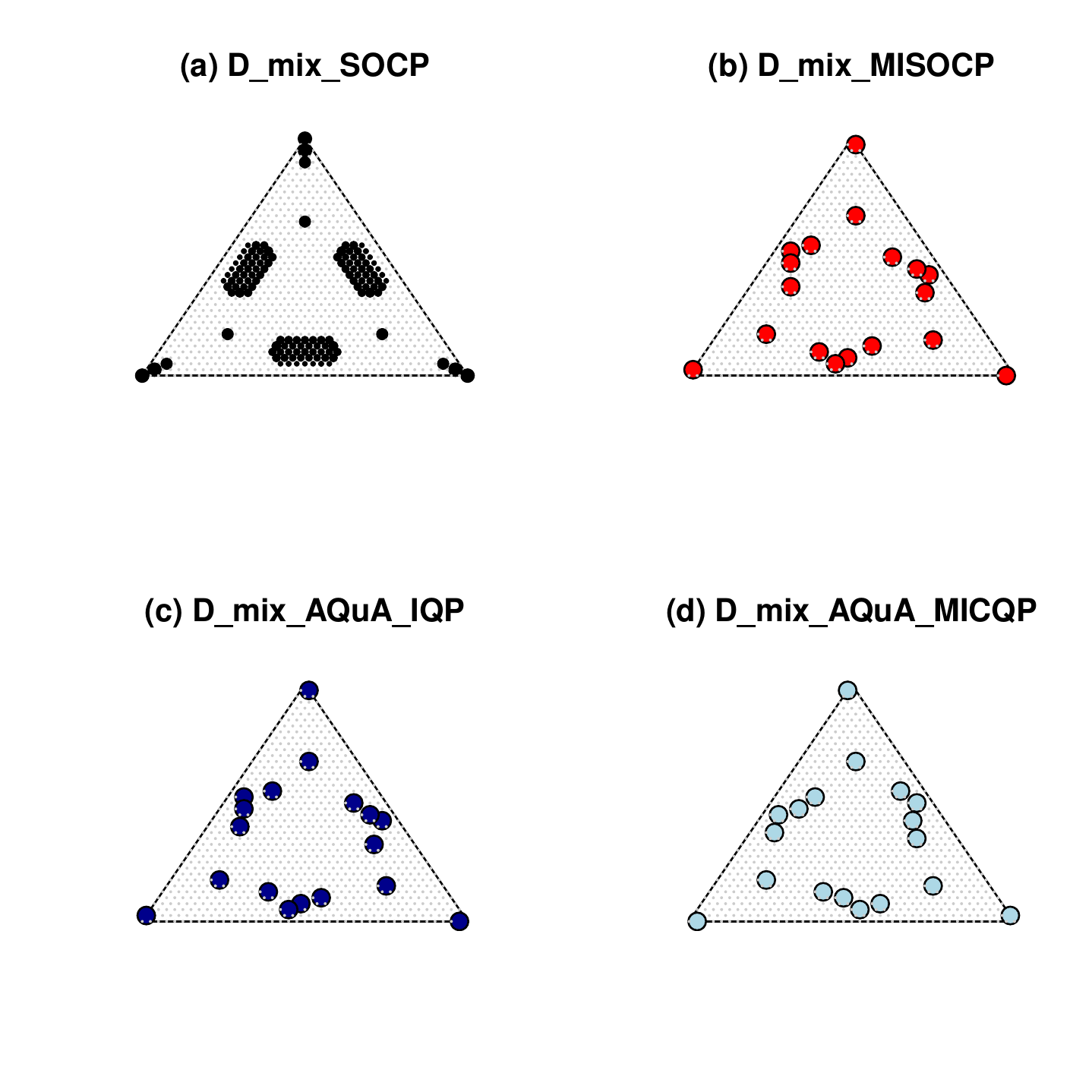}
	\caption{(a) $D$-optimal approximate design for the Scheff\'{e} mixture model \eqref{scheffe} as obtained by the SOCP solver. (b) $D$-efficient exact designs obtained by the MISOCP solver. (c) $D$-efficient exact designs obtained by AQuA via the IQP solver. (d) $D$-efficient exact designs obtained by AQuA with the MICQP solver. The vertices correspond to the pure mixtures $(1,0,0)$, $(0,1,0)$ and $(0,0,1)$, the gray dots represent the discrete design space, with the larger colored dots denoting the obtained designs.}\label{F:scheffeD}
\end{center}
\end{figure}

\begin{figure}[!h]
\begin{center}
\includegraphics[width=\linewidth]{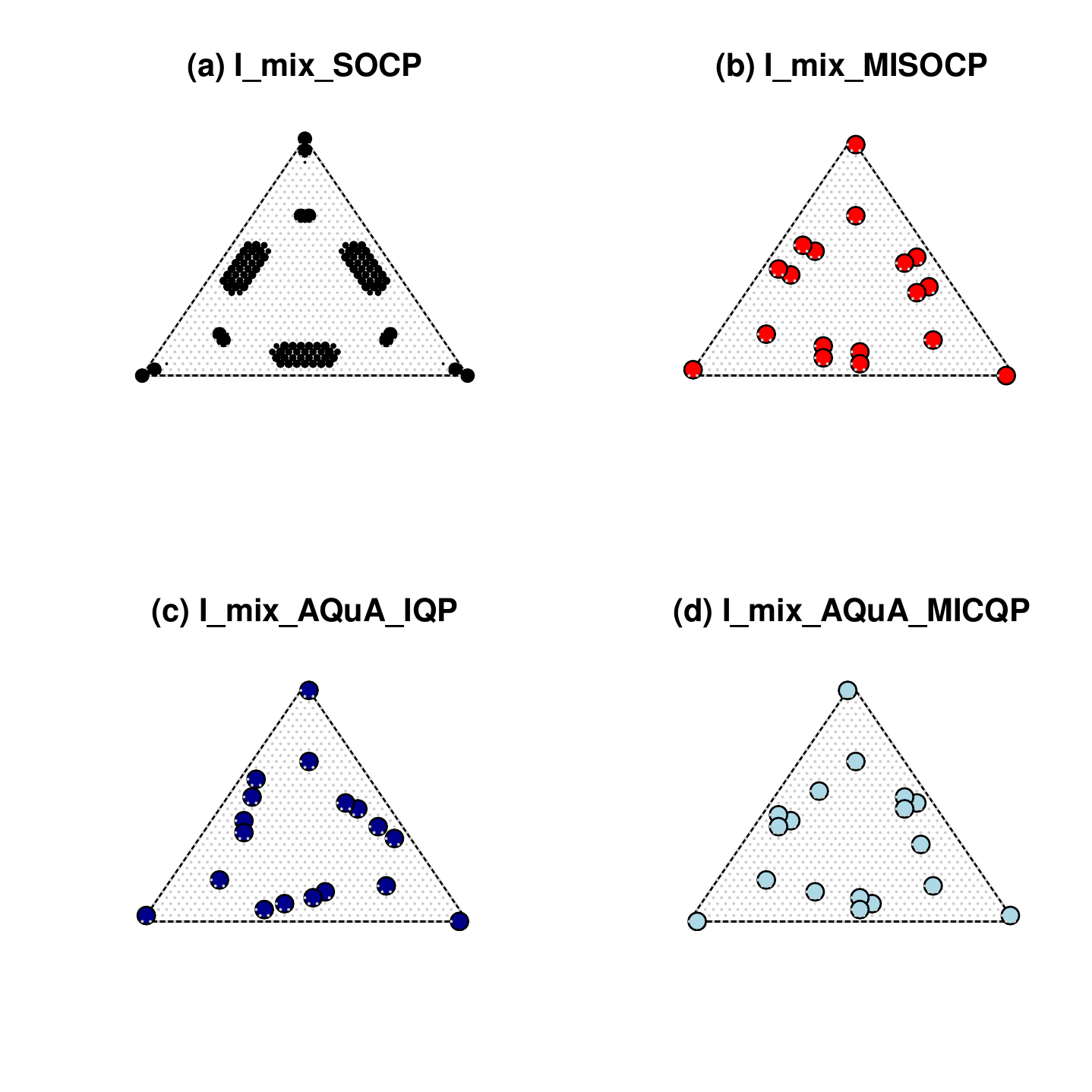}
	\caption{(a) $I$-optimal approximate design for the Scheff\'{e} mixture model \eqref{scheffe} as obtained by the SOCP solver. (b) $I$-efficient exact designs obtained by the MISOCP solver. (c) $I$-efficient exact designs obtained by AQuA via the IQP solver. (d) $I$-efficient exact designs obtained by AQuA with the MICQP solver. The vertices correspond to the pure mixtures $(1,0,0)$, $(0,1,0)$ and $(0,0,1)$, the gray dots represent the discrete design space, with the larger colored dots denoting the obtained designs.}\label{F:scheffeI}
\end{center}
\end{figure}

In this setting, we computed the $D-$, and $I-$optimal exact designs with the MISOCP approach of \citet{SagnolHarman} and with AQuA, realized by both the standard IQP solver and by the MICQP as proposed in Subsection \ref{Sec:MICQP}. The results are depicted in Figures \ref{F:scheffeD} and \ref{F:scheffeI} and described in Tables \ref{T:scheffeD} and \ref{T:scheffeI}. 
\bigskip

We see that all three methods of computing EDs provide designs of similar efficiency, but the conic reformulation of AQuA can decrease the computation time by as much as two orders of magnitude.  
\bigskip

Note that in this case, ER method cannot be used at all to transform AD to ED. Besides the support of ADs in this model being very large, as can be seen in Figures \ref{F:scheffeD} and \ref{F:scheffeI}, the marginal and symmetry constraints cannot be incorporated into ER without its significant modification. 

\begin{table}[!h]
\begin{center}
\begin{tabular}{ c | c | c | c | c }
  \hline
   Panel & type & method & efficiency & time \\
   \hline		
  (a) & appr. & SH-SOCP & $1.000$ & $7.59$ \\
  (b) & exact & SH-MISOCP & $0.98377$ & $616.91$ \\
  (c) & exact & AQuA-IQP & $0.98373$ & $625.53$ \\
  (d) & exact & AQuA-MICQP & $0.98374$ & $8.56$ \\
  \hline  
\end{tabular}
\caption{$D$-optimality, related to Fig. \ref{F:scheffeD}. Approximate and exact designs of the mixture experiment analyzed in Subsection \ref{ss:scheffe} computed by methods SH-SOCP, SH-MISOCP (both described in \citet{SagnolHarman}), AQuA-IQP based on the direct use of an integer quadratic solver and AQuA-MICQP based on the low-rank reformulation in \ref{Sec:MICQP}. The efficiency is computed relative to the optimal approximate design.}\label{T:scheffeD}
\end{center}
\end{table}

\begin{table}[!h]
\begin{center}
\begin{tabular}{ c | c | c | c | c }
  \hline
   Panel & type & method & efficiency & time\\
   \hline		
  (a) & appr. & SH-SOCP & $1.000$ & $3.87$ \\
  (b) & exact & SH-MISOCP & $0.99627$ & $602.96$ \\
  (c) & exact & AQuA-IQP & $0.98928$ & $636.16$ \\
  (d) & exact & AQuA-MICQP & $0.99647$ & $4.54$ \\
  \hline  
\end{tabular}
\caption{$I$-optimality, related to Fig. \ref{F:scheffeI}. Approximate and exact designs of the mixture experiment analyzed in Subsection \ref{ss:scheffe} computed by methods SH-SOCP, SH-MISOCP (both described in \citet{SagnolHarman}), AQuA-IQP based on the direct use of an integer quadratic solver and AQuA-MICQP based on the low-rank reformulation in \ref{Sec:MICQP}. The efficiency is computed relative to the optimal approximate design.}\label{T:scheffeI}
\end{center}
\end{table}

\subsection{$D$- and $I$-optimal subsampling of a dataset under an upper constraint on budget and lower constraint on average quality}\label{ss:wine}

Lastly, we will show that the conic specification of AQuA can be used as a tool for computing EDs for large design spaces, in particular for a constrained information-based subsampling of ``tall'' datasets; see, e.g., \citet{iboss} for a justification of this approach. Here, the purpose is to select a subsample for a screening with the quality based on a linear regression model. In contrast to the existing information-based subsampling methods, we can require a subsample that keeps limits on the numbers of selected objects within given strata, and, simultaneously, a lower constraint on the quality as well as an upper constraints on the price of the subsample. 
\bigskip

To this end we used the wine datafile \citet{wine} that contains data on approximately $150000$ wine reviews from WineEnthusiast. The aim is to subsample this database for a survey, marketing or educational purposes.
\bigskip

After cleaning duplicities and incomplete entries, we were left with $n=111534$ wine reviews containing variables on country of origin, description, points, price, province, title, variety, and winery. Out of this dataset, we are to sample wines so that the upper bound on the combined price of the wines is 1000\$, lower bound on the average quality points is $90$ and there is exactly one wine from each of the $42$ countries. To avoid selecting the same wine more than once, we added the upper bounds $\xi_i \leq 1$ for all $i=1,\ldots,n$. The model used was the linear regression with $m=3$ parameters with intercept, the quality points and the logarithm of the price as explanatory variables.
\bigskip

We will use the robustness of AQuA with respect to the selection of the anchor matrix and the iterative approach explained in Subsection \ref{iter}. For the computation of the first AD, we used the SOCP formulation from \citet{SagnolHarman} applied to a random sub-selection comprising 1500 data-points. Then, we sequentially applied AQuA until convergence.
\bigskip

\begin{figure}[!h]
\begin{center}
\includegraphics[width=\linewidth]{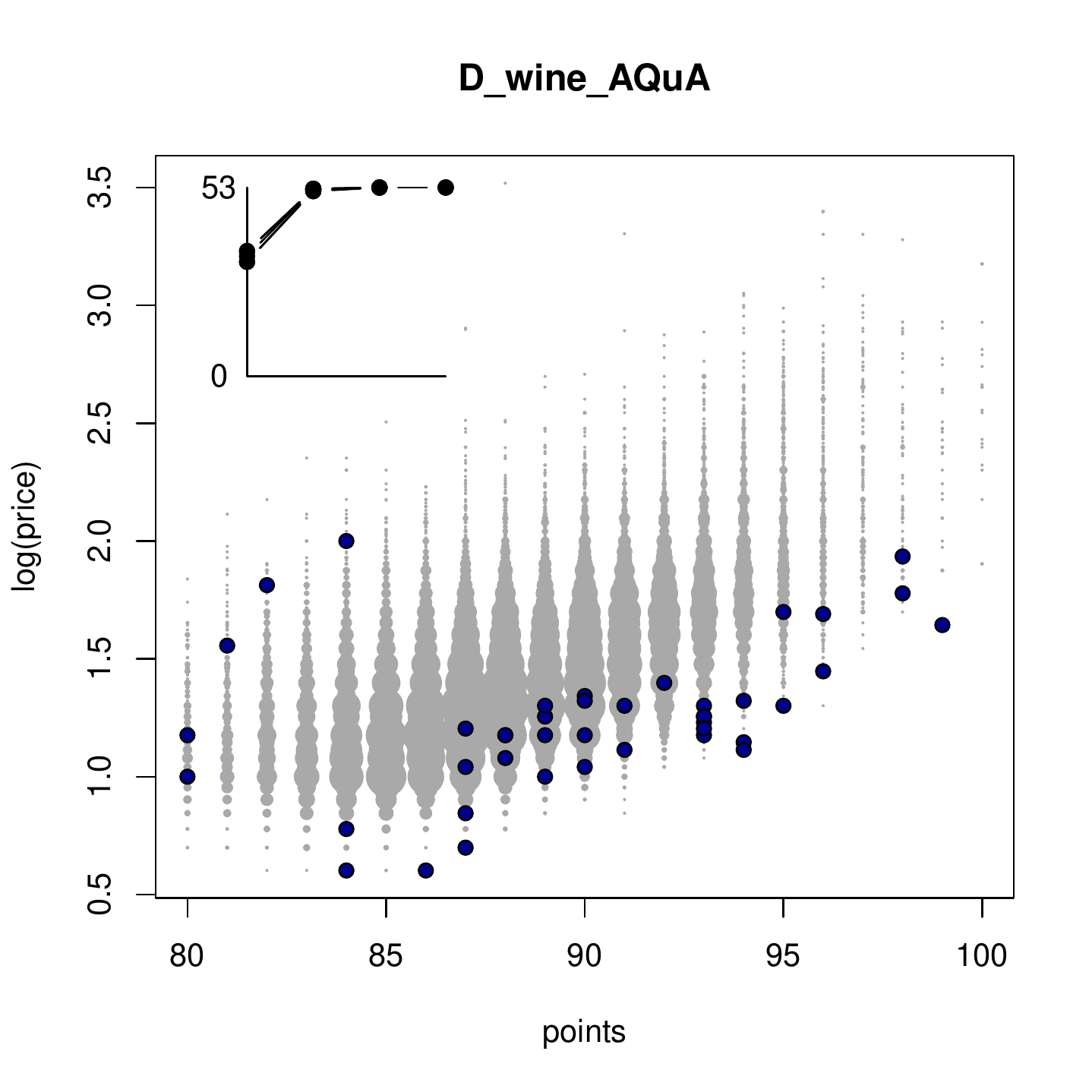}
	\caption{The subsample (blue), chosen from the wine reviews data, based on $D$-criterion. The area of the gray dots is proportional to the density of the full dataset. The inlay shows the convergence of the iterations of the sequential computation of the subsample (the vertical axis is the value of the $D$-criterion).}\label{wineD}
\end{center}
\end{figure}

\begin{figure}[!h]
\begin{center}
\includegraphics[width=\linewidth]{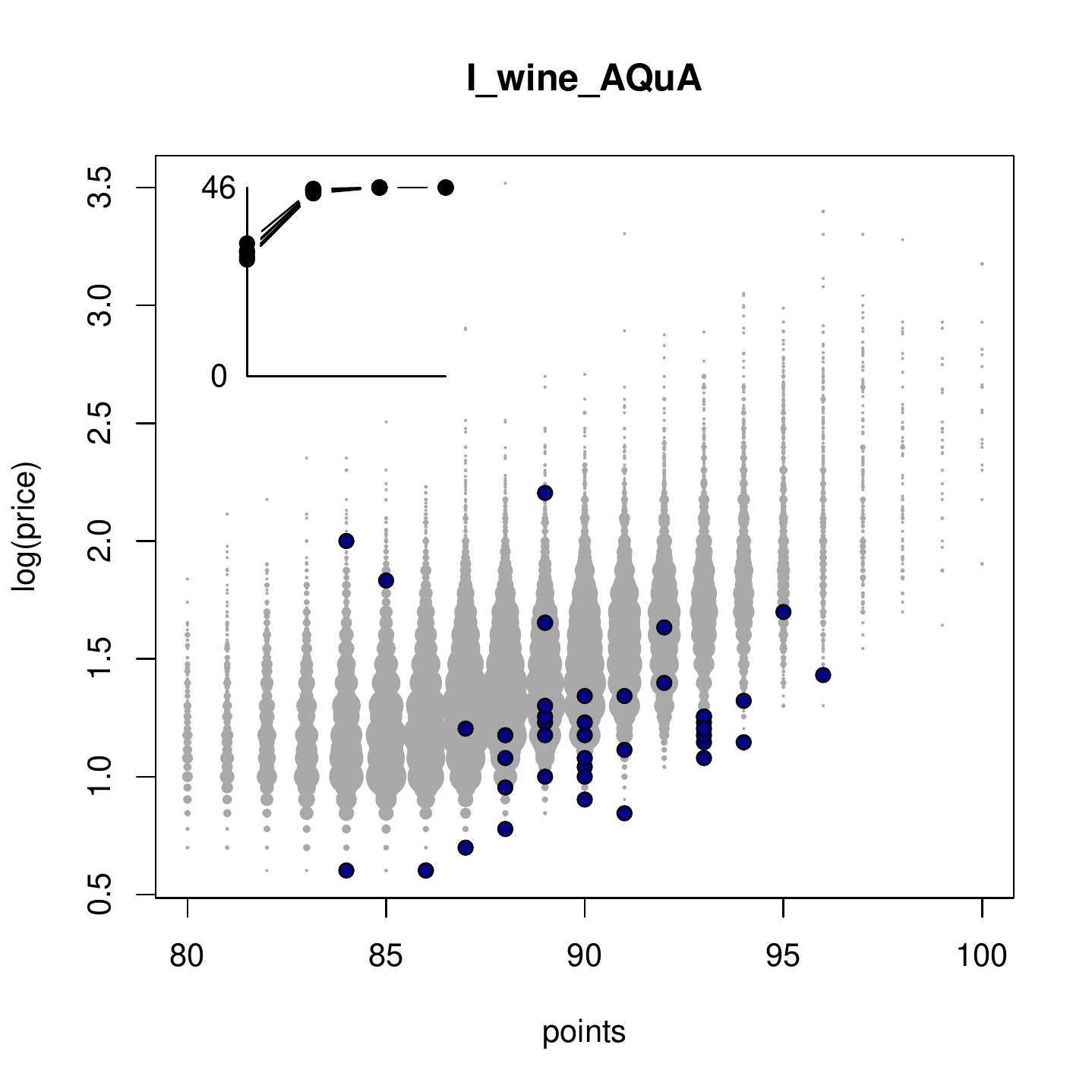}
	\caption{The subsample (blue), chosen from the wine reviews data, based on $I$-criterion. The area of the gray dots is proportional to the density of the full dataset. The inlay shows the convergence of the iterations of the sequential computation of the subsample (the vertical axis is the value of the $I$-criterion).}\label{wineI}
\end{center}
\end{figure}

For both criteria, we run the randomly initiated computation $5$ times, and in every case it converged to the same solution in as few as $4$ steps (including the first, SOCP computation), each taking less than $4$ seconds. The resulting $42$-element subsamples are visualized in Figures \ref{wineD} and \ref{wineI}. It turns out that for both resulting subsamples, the cumulative price of the $42$ wines is exactly 1000\$ and the average quality is exactly $90$ points. To meet the restrictions, the subsample computed using the $D$-optimality criterion is automatically concentrated largely in the region of inexpensive wines of a good quality while still making the samples diverse enough to permit precise estimation of parameter of the linear model. The result based on $I$-optimality is similar, but since $I$-optimality minimizes the average variance throughout all points, the subsample is more concentrated in the area that is most densely populated.   
\bigskip

We also remark that, using a standard computer, AQuA based on the IQP solver (unlike the specific MICQP formulation of AQuA) cannot be applied to design spaces of size larger than a few thousands, because of the quadratic memory requirements. That is, here we again demonstrated the advantage of the proposed conic AQuA approach over the approach of AQuA from \citet{HF}, not only over methods directly based on a MISOCP formulation of the problem as in \citet{SagnolHarman}.

\section{Conclusions}

  We extended the quadratic approximation from a single version of the $D$-criterion used in \citet{HF} to two versions of all Kiefer's criteria with an integer parameter, including the criterion of $A$-optimality and, via transformation, to the criterion of $I$-optimality.
  \bigskip
  
  Importantly, we also proved a low-rank property of the associated quadratic forms and used it to construct efficient conic formulation of the integer quadratic programming problem. The formulation permits using the method of AQuA in case of large design spaces that are out of reach of the previous methods. On the other hand, for smaller design spaces (provided that $m \ll n$ is still satisfied), the conic formulation of AQuA can significantly speed up the computation; in particular, it can rapidly identify an optimal exact design in case where one of the optimal approximate designs is also optimal exact. Moreover, using AQuA it is possible to obtain efficient exact designs for situations with simultaneous constraints on various characteristics of the experiment, e.g., its form, cost, and quality.
  \bigskip
  
  The basic AQuA approach presumes the knowledge of the optimal approximate information matrix. However, because the algorithms for computing optimal approximate designs are well developed and fast, this is not considered to be a drawback. Moreover, there is a large body of literature that provides theorems that explicitly yield optimal approximate designs. Note that with rounding procedures alone, the practical value of optimal approximate designs is weaker since direct heuristic computational methods can often find better designs, entirely circumventing approximate design theory and computation. We prove that optimal approximate designs carry more useful information for the construction of exact designs than is utilized by rounding procedures.
  \bigskip
  
  We also showed that the AQuA approach is generally robust with respect to the misspecification of the optimal information matrix, and can even be used sequentially, starting from an anchor matrix that is far from the approximate optimum. 
\bigskip

  Finally, the approach of AQuA can be extended to various criteria other than those analyzed in this paper; what is needed is only their quadratic approximation\footnote{For the application of the conic improvement, the quadratic forms must have low ranks.}, which can be found either analytically or numerically. This opens up new possibilities for the computation of optimal experimental designs with respect to criteria that are difficult to evaluate.
\bigskip

\textbf{Acknowledgments} The work was supported by Grant No 1/0341/19 from the Slovak Scientific Grant Agency (VEGA).



\end{document}